\documentstyle[prd,aps,eqsecnum,preprint,tighten,floats]{revtex}
\begin{document}
\draft
\input BoxedEPS
\SetTexturesEPSFSpecial
\HideDisplacementBoxes

\preprint{\vbox{\hfill OHSTPY--HEP--TH--95--034\\
          \vbox{\vskip0.5in} }}

\title{Dirac's Legacy: Light-Cone Quantization}

\author{Stephen S. Pinsky}

\address{Department of Physics, The Ohio State University, Columbus,
OH 43210}


\maketitle
\begin{center}
Talk presented at the\\
International Conference on Orbis Scientiae 1996\\
Miami Beach Florida- January 25-28, 1996
\end{center}
\begin{abstract}
In recent years light-cone quantization of quantum field theory has emerged
as a promising method for solving problems in the strong coupling regime.
This approach has a number of unique features that make it
particularly appealing, most notably, the ground state of the free theory
is also a ground state of the full theory. 
\end{abstract}

\newpage
\section{Introduction}

One of the central problems in particle physics is to determine the structure of
hadrons such as the proton and neutron in terms of their fundamental quark and gluon
degrees of freedom.   Over the past twenty years two fundamentally different pictures
of hadronic matter have developed. One, the constituent quark model (CQM), or the
quark parton model is closely related to experimental observation.  The other, 
quantum chromodynamics (QCD) is based on an elegant non-abelian quantum field theory.
The light-front formulation of QCD appears to be  the only practical hope of
reconciling QCD with the CQM. This elegant approach to quantum field theory is a
Hamiltonian gauge fixed formulation that avoids many of the most difficult problems
in the equal time formulation of the theory.  The idea of deriving a null-plane
constituent model from QCD actually dates from the early seventies, and there is a
rich literature on the subject \cite{des73,cat76,ida75a}.  The main thrust of this
talk will be to discuss the complexities of vacuum that are unique to the
light-front formulation of field theories.

An intuitive approach  for solving relativistic bound-state problems would be to
solve the gauge fixed, Hamiltonian eigenvalue problem.  One imagines that there is an
expansion in multi-particle occupation number Fock states. It is clearly  a
formidable task  to calculate the structure of hadrons in terms of their fundamental
degrees of freedom in QCD. Even in the case of  abelian quantum electrodynamics, very
little is known about the nature of the bound state solutions in the strong-coupling
domain. A calculation of bound state structure in QCD has to deal with many
complicated aspects of the theory simultaneously: confinement, vacuum structure,
spontaneous  breaking of chiral symmetry (for massless quarks), while describing a
relativistic many-body system which apparently has unbounded particle number. The
analytic problem of describing QCD bound states is compounded not only by the physics
of confinement, but also by the fact that the wavefunction of a composite of
relativistic constituents has to describe systems of an arbitrary number of quanta
with arbitrary momenta and helicities.  The conventional Fock state expansion based
on equal-time quantization quickly becomes intractable because of the complexity of
the vacuum in a relativistic quantum field theory. Furthermore, boosting such a
wavefunction from the hadron's rest frame to a moving frame is as complex a problem
as solving the bound state problem itself.

Fortunately, ``light-front'' quantization, which can be formulated independent of the
Lorentz frame, offers an elegant avenue of escape.  There are, in fact, many reasons to
quantize relativistic field theories at fixed light-front time. Dirac \cite{dir49}, in
1949, showed that a maximum number of Poincar\'e generators become independent of the
dynamics in the ``front form'' formulation, including certain Lorentz boosts. Unlike the   
equal-time Hamiltonian formalism, quantization on a plane tangential to the light-front can
be formulated without reference to the choice of a specific Lorentz frame. The eigen
solutions of the light-front Hamiltonian have Lorentz scalars $M^2$ as eigenvalues, and
describe bound states of arbitrary four-momentum and invariant mass $M$, allowing  the
computation of scattering amplitudes and other dynamical quantities. 

However, the most remarkable feature of this formalism is the apparent simplicity of the
light-front vacuum. In many theories the vacuum state of the free Hamiltonian is an
eigenstate of the total light-front Hamiltonian. This means that all constituents in a
physical eigenstate  are directly related to that state, and not disconnected vacuum
fluctuations.The Fock expansion constructed on this vacuum state  provides a complete
relativistic many-particle basis for diagonalizing the full theory .The natural gauge for
light-cone Hamiltonian theories is the light-cone gauge $A^+=0$.  In this physical gauge the
gluons have only two physical transverse degrees of freedom, and thus it is well matched to
perturbative QCD calculations. The simplicity of the light-cone Fock representation
relative to that in equal-time quantization arises from the fact that the physical vacuum
state has a much simpler structure on the light cone.  Indeed, kinematical arguments
suggest that the light-cone Fock vacuum is the physical vacuum state.   

The success of the CQM or the Feynman parton model is a powerful for a light-front
formulation of QCD.  The ideas of the parton model seem more easily formulated in the
light-front picture of quantum field theory than in the equal-time formulation. This is a
highly desirable feature if one wishes to have a constituent picture of relativistic bound
states and describe, for example, a baryon as primarily a three-quark state plus a few
higher Fock states {\it \'a la} Tamm and Dancoff.  

Studies of model light-front field theories have shown that the zero modes can in fact
support certain kinds of vacuum structure. The long range phenomena of spontaneous
symmetry breaking \cite{hek91b,bep93,piv94,hip95,rob93} as well as the
topological structure
\cite{kap94,pin94} can in fact be reproduced with a careful treatment of the zero 
mode(s) of the fields in a quantum field theory defined in a finite spatial volume and
quantized at equal light-front time. These phenomena are realized in quite different
ways.  For example, spontaneous breaking of $Z_2$ symmetry in $\phi^4_{1+1}$ occurs
via a {\em constrained} zero mode of the scalar field \cite{bep93,hek92a,rob93}. 
There the zero mode satisfies a nonlinear  equation that relates it to the
dynamical modes in the problem.  At the critical coupling a bifurcation of the
solution occurs
\cite{hek92a,rob93,bep93}.  One must choose one solution to use in formulating the
theory.  This choice is analogous to what in the conventional language one would call
the choice of vacuum state. These solutions lead to new operators in the Hamiltonian
which break the $Z_2$ symmetry at and beyond the critical coupling. The various
solutions contain $c$-number pieces which produce the possible vacuum expectation
values of $\phi$. The properties of the strong-coupling phase transition in this model
are reproduced, including its second-order nature and a reasonable value for the
critical coupling\cite{bep93,piv94}. 

Apart from the
question of whether or not VEVs arise, solving the constraint equations really
amounts to determining the Hamiltonian (and other Poincar\'e generators).  In
general, $P^-$ becomes very complicated when the zero mode contributions are
included; this is in some sense the price one pays to achieve a formulation with a
simple vacuum. It may be possible to think of the discretization as a cutoff which
removes states with $0<p^+<\pi/L,$ and the zero mode contributions to the Hamiltonian
as effective interactions that restore the discarded physics. In the light-cone power
counting analysis of Wilson \cite{wiw94} it is clear that there will be a huge number of
allowed operators. 

Quite separately, a {\em dynamical} zero mode was shown in Ref. \cite{kap94} to arise
in pure $SU(2)$ Yang-Mills theory in 1+1 dimensions.  A complete fixing of the gauge
leaves the theory with one degree of freedom, the zero mode of the vector potential
$A^+$. The theory has a discrete spectrum of zero-$P^+$ states corresponding to modes
of the flux loop around the finite space.  Only one state has a zero eigenvalue of
the energy $P^-$, and is the true ground state of the theory. The nonzero eigenvalues
are proportional to the length of the spatial box, consistent with the flux loop
picture.  This is a direct result of the topology of the space. As the theory
considered there was a purely topological field theory the exact solution was
identical to that in the conventional equal-time approach on the analogous spatial
topology \cite{het93,las92,raj92}.  In the
present context, the difficulty is that the zero mode in $A^+$ is in fact gauge-invariant,
so that the light-cone gauge $A^+=0$ cannot be reached.  Thus one has a pair of
interconnected problems: first, a practical choice of gauge; and second, the presence
of constrained zero modes of the gauge field.  In several recent papers
\cite{kap93b,kap94,pik96} these problems were separated and consistent gauge
fixing conditions were introduced to allow isolation of the dynamical and constrained fields. 

The study of these low dimensional theories is part of a long-term program to attack
QCD$_{3+1}$ through the zero mode sectors starting with studies of lower dimensional
theories which are themselves zero mode sectors of higher dimensional theories
\cite{kap93b,kap94}.  A complete gauge fixing has recently been given for
discrete light-cone quantized QED$_{3+1}$ which further supports this program
\cite{kar94} and one sees how zero modes naturally arise and the special role that
they play.  In appears that the central problem in light-front QCD will be to disentangle
and solve the constraints for the dependent zero modes in terms of the independent
fields in the context of a particular gauge fixing.

\subsection {Constrained Zero Modes}
 
As mentioned previously, the light-front vacuum state is simple; it contains no
particles in a massive theory.  However, one commonly associates important long range
properties of a field theory with the vacuum:  spontaneous symmetry breaking, the Goldstone
pion, and color confinement.  How do these complicated phenomena manifest
themselves in light-front field theory?  

If one cannot associate long range phenomena with the vacuum state itself, then
the only alternative is the zero momentum components or ``zero modes'' of the
field (long range $\leftrightarrow$ zero momentum).  In some cases, the zero mode
operator is not an independent degree of freedom but obeys a constraint equation. 
Consequently, it is a complicated operator-valued function of all the other modes
of the field~\cite{may76}.
This problem has recently been attacked from several directions. The question of
whether boundary conditions can be consistently defined in  light-front
quantization has been discussed by McCartor and Robertson~\cite{mcr92} 
and Lenz~\cite{len90,let91}. They have shown that for massive theories the energy and
momentum derived from light-front quantization are conserved and are equivalent to the
energy and momentum one would normally write down in an equal-time theory. Heinzl and
Werner {\em et al.}~\cite{hek92a,hek91b}  considered 
$\phi^4$ theory in (1+1)--dimensions and solved the zero mode constraint equation by
truncating the equation to one  particle and retaining all modes.   They
implicitly retain a two particle contribution  in order to obtain finite results.
Other authors~\cite{hav87} find that, for  theories  allowing spontaneous symmetry
breaking, there is a degeneracy  of light-front vacua and the true vacuum state can differ
from the perturbative vacuum through the addition of  zero mode quanta.   In addition to
these approaches there are many  others~\cite{bur93}.

The definitive analysis by Pinsky, van de Sande, Bender and Hiller \cite{bep93,vap95,hip95} of
the zero mode constraint  equation for (1+1)--dimensional $\phi^4$ field theory 
with symmetric  boundary conditions shows how spontaneous 
symmetry breaking occurs within the context of this model.  This theory has a $Z_2$ symmetry
$\phi \rightarrow - \phi$ which is spontaneously broken for some values of the mass and
coupling.    Their approach is to
apply a Tamm-Dancoff truncation to the Fock space.  Thus operators are finite matrices and
the operator valued constraint equation for the zero mode can be solved numerically. The
truncation assumes that states  with a large number of particles or large  momentum do not
have an important  contribution to the zero mode.

\begin{figure}
\centering\BoxedEPSF{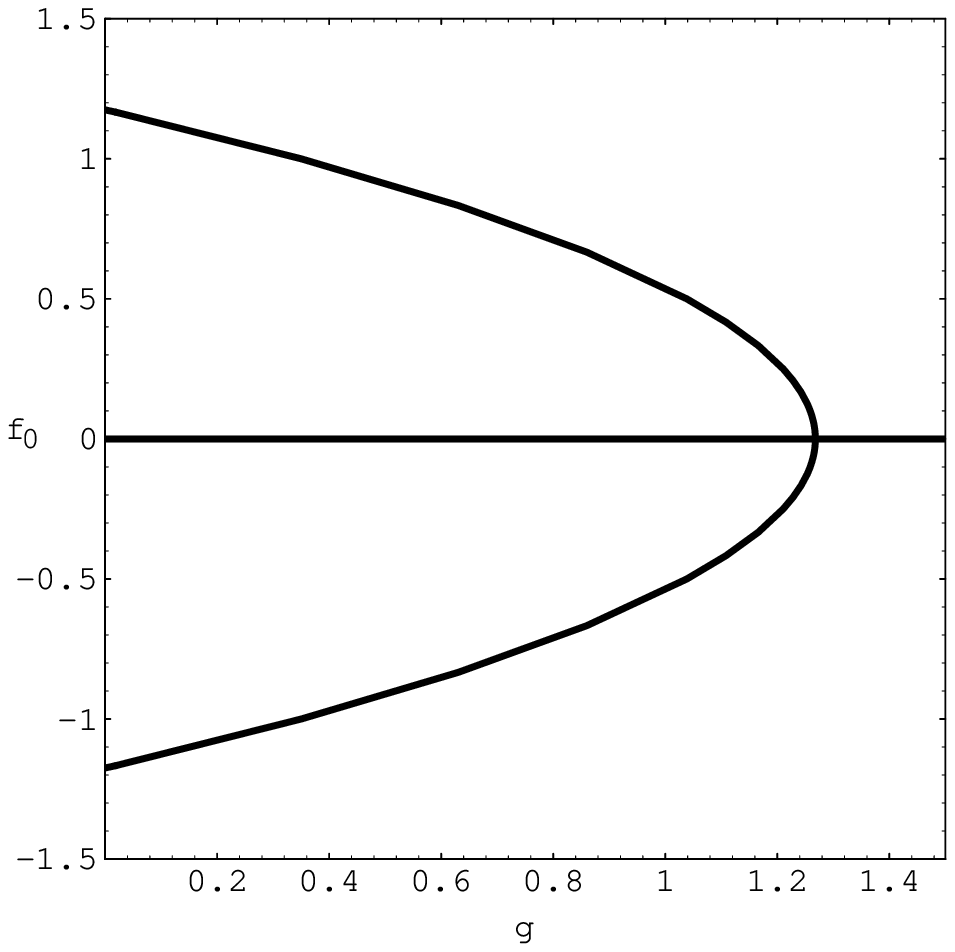 scaled 500}
\caption{\label{ff1}
$f_0 = \protect\sqrt {4 \pi} \langle 0 | \phi | 0 \rangle$ vs.\ $g = 24 \pi
\mu^2/\lambda$ in the one mode case with $N=10$.}
\end{figure}

One finds the following general behavior:  for small coupling (large $g$,
where\linebreak[4]\ $g \propto 1/{\rm coupling}$) the constraint equation has a
single solution and the field has no vacuum expectation value (VEV). As one
increase the coupling (decrease $g$) to  the ``critical
coupling''  $g_{\rm critical}$, two  additional solutions which give the field a
nonzero VEV appear.   These solutions differ only infinitesimally from the first
solution near the  critical coupling, indicating  the presence of a second order
phase transition. Above the critical coupling ($g < g_{\rm critical}$), there are
three solutions:  one with zero VEV, the ``unbroken phase,'' and two with nonzero
VEV, the ``broken phase'' .   The ``critical curves'' shown in Figure~\ref{ff1}, is a
plot the VEV as a function of $g$.  
%
%


Since the vacuum in this theory is trivial, all of the long range  properties 
must occur in the operator structure of the Hamiltonian. Above the critical
coupling ($g < g_{\rm critical}$) quantum oscillations spontaneously break the 
$Z_2$ symmetry of the theory. In a loose analogy with a symmetric  double well
potential,  one has two new Hamiltonians for the  broken phase, each producing
states localized in one of the wells.   The structure of the two Hamiltonians is
determined from the broken phase  solutions of the zero mode constraint
equation.   One finds that the two Hamiltonians have  equivalent spectra.   In a
discrete theory without zero modes it is well known that, if one  increases  the
coupling sufficiently, quantum correction will generate tachyons  causing the 
theory to break down near the critical coupling.   Here the zero mode generates
new interactions that prevent  tachyons from developing.   In effect what happens
is that, while  quantum  corrections attempt to drive the mass negative, they also
change the  vacuum energy  through the zero mode and the diving mass eigenvalue
can never catch the  vacuum eigenvalue.   Thus, tachyons never appear in the
spectra.

In the weak coupling limit ($g$ large) the  solution to the constraint equation
can be obtained in perturbation theory.  This solution does not break the $Z_2$
symmetry and is believed to simply insert the missing zero momentum contributions
into  internal propagators.  This must  happen if light-front perturbation theory
is to agree with equal-time perturbation  theory.  

Another way to investigate the zero mode is to study the spectrum of the field
operator  $\phi$. Here one finds a picture that agrees with the symmetric double
well potential analogy.  In the broken phase, the field is localized in one of the
minima of the  potential and there is tunneling to the other minimum.

\subsection{Canonical Quantization}

The details of the Dirac-Bergmann prescription and its application to  the
system considered here are discussed elsewhere in the  literature
\cite{may76,wit89}. For a classical field the $(\phi^4)_{1+1}$ 
Lagrange density is
\begin{equation} {\cal L} = \partial_+\phi\partial_-\phi - {{\mu^2}\over 2} \phi^2
-  {\lambda
\over 4!} \phi^4\;. 
\end{equation}
One puts the system in a box of length $d$ and impose periodic  boundary
conditions.   Then
\begin{equation}
\phi\!\left(x\right) = {1\over{\sqrt d}} \sum_n q_n(x^+) e^{i k_n^+ x^-}\; , 
\end{equation}
where $k_n^+ =2\pi n/d$ and summations run over all integers unless otherwise
noted.

The $\int dx^- \, \phi(x)^n$ minus the zero mode part is 

\begin{equation}
\Sigma_n = {1\over{n!}} \sum_{i_1, i_2, \ldots, i_n \neq 0} q_{i_1}  q_{i_2}
\ldots q_{i_n}\, \delta_{i_1 + i_2 + \ldots + i_n, 0} 
\end{equation}
and the canonical Hamiltonian is
\begin{equation} P^- = {{\mu^2 q_0^2}\over 2} + \mu^2 \Sigma_2 + {{\lambda 
q_0^4}\over{4! d}}
 + {{\lambda q_0^2 \Sigma_2}\over{2! d}} + {{\lambda q_0 \Sigma_3}\over  d} +
{{\lambda \Sigma_4}\over d}\;. 
\end{equation}
Following the Dirac-Bergmann prescription, one identify  first-class constraints
which define the conjugate momenta
\begin{equation} 0 = p_n - i k_n^+ q_{-n}\;, 
\end{equation}
where
\begin{equation}
\left[q_m,p_n\right] = {{\delta_{n, m}}\over 2}\;,\quad m ,n \neq 0\;. 
\end{equation}
The secondary constraint is,
\begin{equation} 0 = \mu^2 q_0 + {{\lambda q_0^3}\over{3! d}} + {{\lambda q_0 
\Sigma_2}\over d} + {{\lambda \Sigma_3}\over d}\; , 
\end{equation}
which determines the zero mode $q_0$. This result can also  be obtained by
integrating the equations of motion. 

To quantize the system one replaces the classical fields with the  corresponding
field operators, and the Dirac bracket by  $i$ times a commutator. One must choose
a regularization and an operator-ordering prescription in order to make the system
well-defined.
 
One begin by defining creation and annihilation operators 
$a_k^\dagger$ and $a_k$,
\begin{equation} q_k = \sqrt {d\over{4 \pi \left| k \right|}} \: a_k\;,
\quad a_k = a_{-k}^\dagger\;,\quad  k\neq 0\; , 
\end{equation}
which satisfy the usual commutation relations
\begin{equation}
\left[a_k,a_l\right] =0\;,
\quad\left[a_k^\dagger,a_l^\dagger\right] =0\;,\quad
\left[a_k,a_l^\dagger\right] =\delta_{k, l}\; ,\quad k,l > 0\; .
\end{equation}
Likewise, one defines the zero mode operator
\begin{equation} 
q_0 = \sqrt{d\over{4 \pi}} \: a_0\;.
\end{equation}
In the quantum case, one normal orders $\Sigma_n$. 

General arguments suggest that the Hamiltonian  should be symmetric
ordered~\cite{bem86}. However, it is not clear how one should treat the 
zero mode since it is not a dynamical field. As an {\em ansatz} one treats
$a_0$ as an ordinary field operator when symmetric ordering the Hamiltonian.  The
tadpoles are removed from the symmetric ordered Hamiltonian by  normal ordering
the terms having no zero mode factors and by subtracting,
\begin{equation} 
{3\over{2}}\; a_0^2\sum_{n\neq 0} {1\over{|n|}} \; .
\end{equation}
In addition, one subtract a constant so that the  VEV of $H$ is zero. Note
that this renormalization prescription is equivalent to a conventional mass
renormalization and  does not introduce any new operators 
into the Hamiltonian. The constraint equation for the zero mode can
be obtained by taking a derivative of $P^-$ with respect to $a_0$. One finds,
\begin{equation} 
0 = g a_0 + a_0^3 +  \sum_{n\neq 0} {1\over{|n|}}
\left( a_0 a_n a_{-n} + a_n a_{-n} a_0 + a_n a_0 a_{-n} -
{3 a_0 \over 2} \right)+ 6 \Sigma_3 \;  
\end{equation}
\noindent where $g= 24 \pi \mu^2/\lambda$. 
It is clear from the general structure of constraint equation that $a_0$ as a
function of the other modes is not necessarily odd under the transform $a_k \to
-a_k$, $k \neq 0$ associated with the $Z_2$ symmetry of the system. Consequently,
the zero mode can induce $Z_2$ symmetry breaking in the Hamiltonian.

In order to render the problem tractable, one can impose a Tamm-Dancoff truncation on
the Fock space. Define $M$ to be the number of nonzero modes and $N$ to be the 
maximum number of allowed particles.  Thus, each state in the truncated Fock space
can be represented by a vector of length $S=\left(M+N\right)!/\left(M! N! \right)$
and  operators  can be represented by $S \times S$ matrices.   One can define the
usual Fock space basis, $\left|n_1,n_2,\ldots,n_M\right\rangle$.
where $n_1 + n_2 + \ldots +n_M \leq N$.  In matrix form, $a_0$ is real and
symmetric. Moreover, it is  block diagonal in states of equal $P^+$ eigenvalue. 

\subsection{ Perturbative Solution of the Constraints}
 
In the limit of large $g$, one can solve the constraint equation
perturbatively  . Then one substitutes the
solution back into the Hamiltonian and calculates various amplitudes to arbitrary
order in $1/g$ using Hamiltonian  perturbation theory.  

It can be shown that the solutions of the constraint equation
and the  resulting Hamiltonian are divergence free to all orders in perturbation
theory for both the broken and unbroken phases.  The perturbative solution for the
unbroken  phase is
\begin{equation}
 a_0 = - \frac{6}{g} \Sigma_3 +  \frac{6}{g^2}\left(2 \Sigma_2
\Sigma_3+ 2
\Sigma_3  \Sigma_2+\sum_{k=1}^M \frac{a_k \Sigma_3 a_k^\dagger + a_k^\dagger
\Sigma_3 a_k -\Sigma_3}{k}\right)  + O(1/g^3) \; .
\label{pzm}
\end{equation}
\noindent Substituting this into the Hamiltonian, one obtains a complicated but well
defined expression.

The finite volume box acts as an infra-red regulator and the  only possible
divergences are ultra-violet.  Using diagrammatic language, any loop  of momentum
$k$ with $\ell$ internal lines has asymptotic form $k^{-\ell}$. Only the case of
tadpoles ${\ell}=1$ is divergent.  If there are multiple loops, the effect is to
put factors of $\ln(k)$ in the numerator and the divergence structure is
unchanged.  Looking at Eq.~(\ref{pzm}), the only possible tadpole is from the
contraction in the term
\begin{equation}
\frac{a_k \Sigma_3 a_{-k}}{k}
\end{equation}
which is canceled by the $\Sigma_3/k$ term.  This happens to all orders in
perturbation theory:  each tadpole has an associated term which cancels it.  
Likewise, in the Hamiltonian one has similar cancellation.  As
with the zero mode, such cancellations occur to all orders in perturbation
theory. 
\begin{figure}
\centering\BoxedEPSF{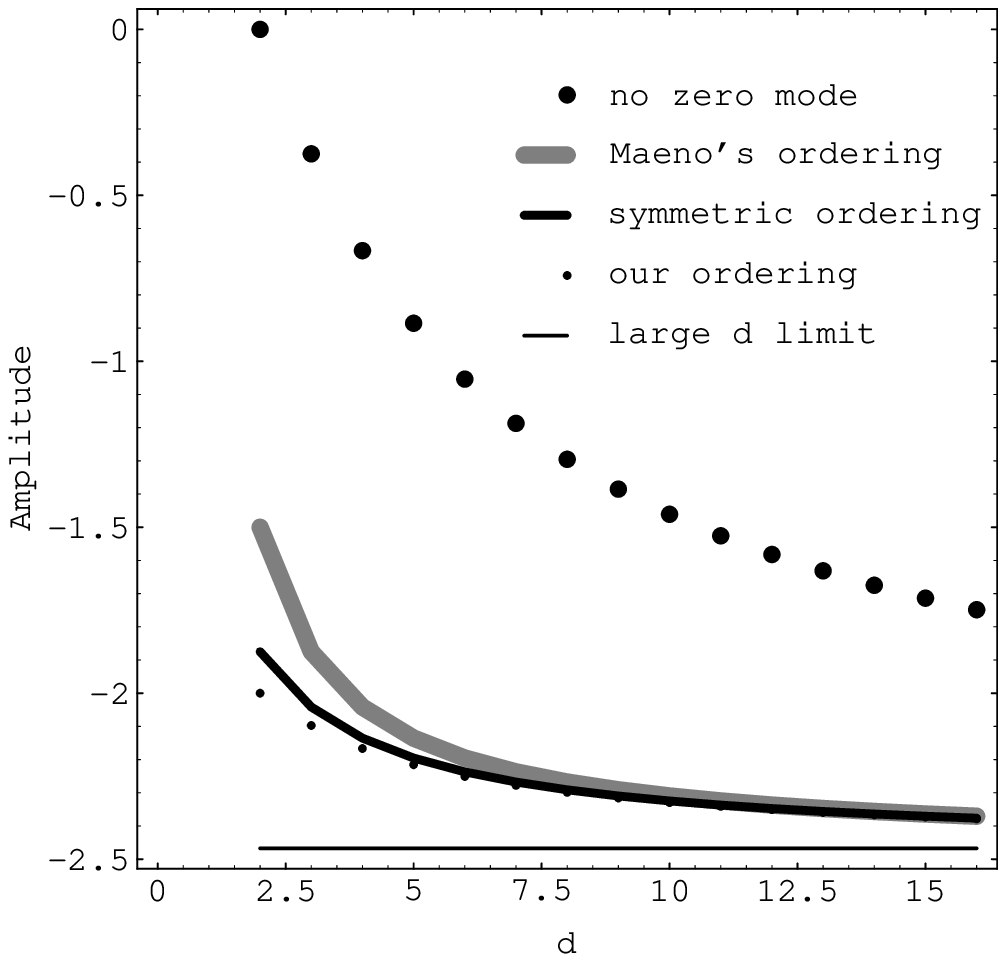 scaled 750}
\caption{\label{ff14} Convergence to the large $d$ limit of $1 \to 1$ setting 
$E=g/p$ and dropping any constant terms. }
\end{figure}
For the unbroken phase, the effect of the zero mode should vanish in the  infinite
volume limit, giving a ``measure zero'' contribution to the continuum Hamiltonian.
However, for finite box volume the zero mode does contribute,  compensating
for the fact that the longest wavelength mode has been  removed from the system.
Thus, inclusion of the zero mode improves convergence to the infinite  volume
limit; it acts as a form of infra-red renormalization. In addition, one can use
the perturbative expansion of the zero mode to study the operator ordering
problem.  One can directly compare our operator ordering ansatz with a truly Weyl
ordered Hamiltonian and with Maeno's operator  ordering ansatz \cite{mae94}.
As an example, let us examine $O(\lambda^2)$ contributions to the  processes $1
\rightarrow 1$ which as 
shown in Figure~\ref{ff14}. including the zero
mode greatly  improves convergence to the large volume limit. The zero mode
compensates, in an optimal manner, for the fact that one  has removed the longest
wavelength mode from the system.

\subsection{Non-Perturbative Solution: One Mode, Many Particles}
 
Consider the case of one mode $M=1$ and many particles.   
In this case, the zero-mode is diagonal and can be written as
\begin{equation} a_0 = f_0 \left| 0 \right\rangle\left\langle 0 \right| +
{\sum_{k=1}^N f_{k} \left| k \right\rangle\left\langle k \right|}\; .
\label{onemode}
\end{equation}
\noindent Note that $a_0$ in (\ref{onemode}) is even under 
$a_k \to -a_k$, $k \neq 0$ and any non-zero solution  breaks the $Z_2$ symmetry of
the original Hamiltonian. The VEV is given by
\begin{equation}
\langle 0 | \phi | 0 \rangle ={1\over{\sqrt{4 \pi}}}
\langle 0| a_0 | 0\rangle = {1\over{\sqrt{4 \pi}}} f_0 \; .
\end{equation}
Substituting (\ref{onemode}) into the constraint equation
and sandwiching the constraint equation between Fock states, one get a recursion
relation for $\left\{f_{n}\right\}$:
\begin{equation} 0 =  g f_{n} + {f_{n}}^{3} + (4n - 1) f_{n} +
\left(n+1\right) f_{n+1} + n f_{n-1} 
\label{recursion}
\end{equation}
%


\noindent where $n \leq N$, and one define $f_{N+1}$ to be unknown. Thus, 
$\left\{ f_{1}, f_{2}, \ldots, f_{N+1} \right\}$  is uniquely determined by a 
given choice of $g$ and $f_0$. In particular, if $f_{0}=0$ all the $f_{k}$'s are
zero independent of $g$. This is the unbroken phase. 

Consider the asymptotic 
behavior for large $n$.  If $f_{n}\gg 1$ in this limit, then the ${f_{n}}^3$ term
will  dominate and
\begin{equation} f_{n+1} \sim {{f_{n}^3}\over n}\;,  
\end{equation}
thus,
\begin{equation} {\lim_{n\to\infty} f_{n}} \sim (-1)^n \exp\!\left(3^n  {\rm
constant}\right)\;.  
\end{equation}
One must reject this rapidly growing solution. 
Hence, one only seek solutions where $f_{n}$ is small for large $n$. 
For large $n$, the terms linear in $n$ 
dominate and  Eq.~(\ref{recursion}) becomes
\begin{equation} f_{n+1} + 4 f_{n} + f_{n-1} = 0\; .
\end{equation}
There are two solutions to this equation:
\begin{equation} f_{n} \propto \left(\sqrt{3} \pm 2\right)^{n}\; .
\end{equation}
One must reject the plus solution because it grows with $n$.    This gives the
condition
\begin{equation} -{{\sqrt{3} - 3 + g}\over{2\sqrt 3}} = K\;,
\quad K= 0, 1, 2\ldots 
\label{criticals}
\end{equation}
%

\noindent Concentrating on the $K=0$ case, one finds a critical coupling
\begin{equation} g_{\rm critical} = 3 - \sqrt{3}
\end{equation}
or
\begin{equation}
\lambda_{\rm critical}=4\pi\left(3+\sqrt{3}\right) \mu^2 \approx 60\mu^2.
\end{equation}
In comparison, values of 
$\lambda_{\rm critical}$ from $ 22 \mu^2 $ to $ 55 \mu^2 $ have  been reported for
equal-time quantized calculations \cite{cha76,abe76,fuk87,krg87}. The solution to the
linearized equation is an approximate solution to the full
Eq.~(\ref{recursion})  for
$f_0$ sufficiently small. Next, one needs to determine solutions of the full
nonlinear equation which converge for large $n$.

One can study the critical curves by looking for numerical solutions to
Eq.~(\ref{recursion}). The method used here is to find values of $f_{0}$ and
$g$ such that $f_{N+1}=0$. Since one seeks a solution where $f_{n}$ is decreasing
with $n$,  this is a good approximation.  One finds that for $g>3-\sqrt 3$ the only
real solution is $f_{n}=0$ for all $n$.  For $g$ less than $3-\sqrt{3}$ there are
two additional solutions.  Near the critical point $\left| f_{0} \right|$ is small
and
\begin{equation} f_{n} \approx f_{0} \left(2 - \sqrt 3\right)^n\; .
\end{equation}
The critical curves are shown in Figure~\ref{ff1}. These solutions converge
quite rapidly with $N$. The critical curve for the broken phase is approximately
parabolic in shape:
\begin{equation} g \approx  3-\sqrt 3 - 0.9177 f_{0}^2 \; . 
\end{equation}

One can also study the eigenvalues of the Hamiltonian for the  one mode case.  The
Hamiltonian is diagonal for this Fock space truncation and,
\begin{equation}
\left\langle n \right| H \left| n \right\rangle = {3\over 2} n (n-1) + n g
-{f_n^4\over 4} - {{2 n+1}\over 4} f_n^2 +{{n+1}\over 4}  f_{n+1}^2 + {n\over 4}
f_{n-1}^2 - C\; .
\label{onehamiltonian}
\end{equation}
\noindent The invariant mass eigenvalues are given by
\begin{equation}
   P^2 | n \rangle =
   2 P^+ P^- | n \rangle = 
\frac{n \lambda  \langle n | H | n \rangle}{24 \pi} | n \rangle
\end{equation}
%
%
\begin{figure}
\centering\BoxedEPSF{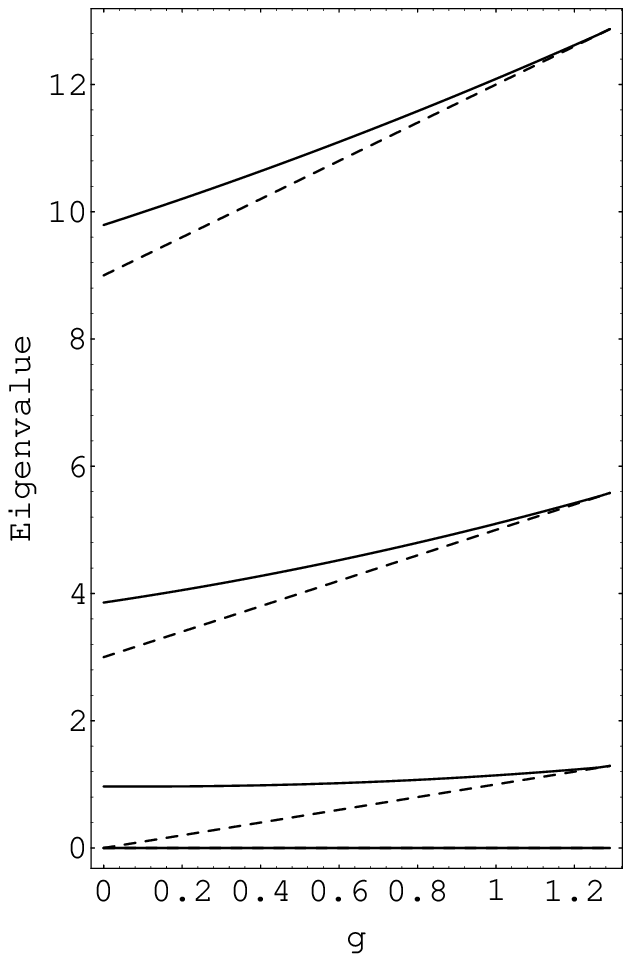}
\caption{\label{ff4} The lowest three energy eigenvalues for the one mode  case as
a function of $g$ from the numerical solution of
Eq.~(\protect\ref{onehamiltonian}) with $N=10$.  The dashed lines are for the
unbroken phase $f_0 = 0$  and the solid lines are for the broken phase $f_0 \neq
0$.}
\end{figure}
%

In Figure~\ref{ff4} the dashed lines show the first few eigenvalues as a function of $g$
without the zero-mode.   When one include the broken phase of the zero mode, the energy
levels shift as shown by the solid curves.  For $g < g_{\rm critical}$  the energy levels
increase  above the value they had without the zero mode.  The higher levels change very 
because $f_n$ is small for large $n$.
In the more general case of many modes and
many particles many of the features that were seen in the one mode and one
particle cases remain.   
  
One can also investigate the shape of the critical curve near the critical
coupling as a function of the cutoff $K$.  In scalar field theory, 
$\langle 0 |\phi|0\rangle$ acts as  the order parameter of the
theory.  Near the critical coupling,  one can fit the VEV to some
power of 
$g-g_{\rm critical}$; this will give us the associated critical exponent $\beta$,
\begin{equation}
\langle 0 | a_0 | 0 \rangle \propto \left(g_{\rm critical}-g\right)^\beta \; .
\end{equation}
They have calculated this as a function of cutoff and found a result consistent
with $\beta=1/2$, independent of cutoff $K$. 
The theory
$\left(\phi^4\right)_{1+1}$ is in the same universality class as the Ising model
in 2 dimensions and the correct critical exponent for this universality class is
$\beta=1/8$. If one were to use the mean field approximation to calculate the
critical exponent,  the  result would be $\beta=1/2$. This is what was
obtained in this calculation.  Usually, the presence of a mean field result
indicates that one is not probing all length scales properly. If one had a cutoff
$K$ large enough to include many length scales, then the critical exponent should
approach the correct value.   However, one cannot be certain that this is the
correct explanation of our result since no evidence that 
 $\beta$ decreases with increase $K$ is seen.

\subsection{Spectrum of the Field Operator}

How does the zero mode affect the field itself? Since $\phi$ is a
Hermitian operator it is an observable of  the system and one can measure
$\phi$ for a given  state $|\alpha\rangle$.   $\tilde{\phi}_i$ and 
$|\chi_i\rangle$ are the eigenvalue and eigenvector respectively of $\sqrt{4 \pi}
\phi\,$:
\begin{equation}
 \sqrt{4 \pi} \phi \, |\chi_i\rangle = 
\tilde{\phi}_i|\chi_i\rangle \;,
\;\;\;\; \langle \chi_i | \chi_j \rangle = \delta_{i,j}\; .
\end{equation}
The expectation value of $\sqrt{4 \pi} \phi$ in the state $|\alpha\rangle$ is
$\sum_i \tilde{\phi}_i \left|\langle \chi_i | \alpha \rangle \right|^2$.  
%
%
\begin{figure}
\centering\BoxedEPSF{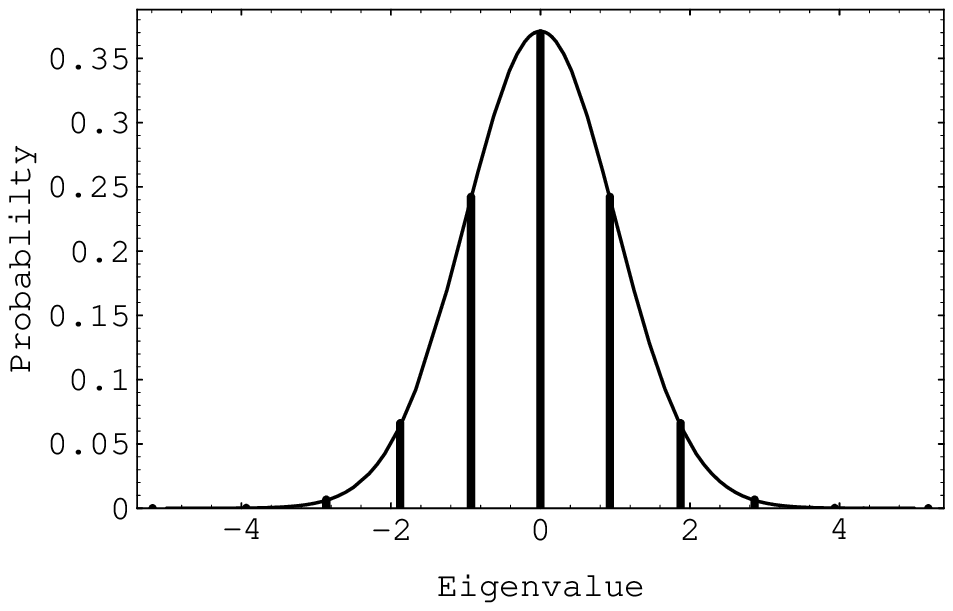}
\caption{
\label{ff10} Probability distribution of eigenvalues  of $\protect\sqrt{4 \pi}
\phi$ for the vacuum with  
$M=1$, $N=10$, and no zero mode.  Also shown is the infinite $N$ limit from
Eq.~(\protect\ref{gaussian}). }
\end{figure}
In the limit of large $N$, the probability distribution becomes continuous.  If
one ignores the zero mode, the probability of obtaining $\tilde{\phi}$ as the result
of a measurement of 
$\sqrt{4 \pi} \phi$ for the vacuum state is
\begin{equation} P\left(\tilde{\phi}\right) = \frac{1}{\sqrt{2 \pi \tau}}\, 
\exp\left(-\frac{\tilde{\phi}^2}{2 \tau}\right)\, d\tilde{\phi} 
\label{gaussian}
\end{equation}
where $\tau = \sum_{k=1}^M 1/k$.  The probability distribution comes  from the 
ground state wave function of the Harmonic oscillator where one identifies $\phi$
with the position operator. This is just the Gaussian fluctuation of a free
field.  
When $N$ is finite, the distribution becomes discrete as  shown in
Figure~\ref{ff10}.
%
%
\begin{figure}
\BoxedEPSF{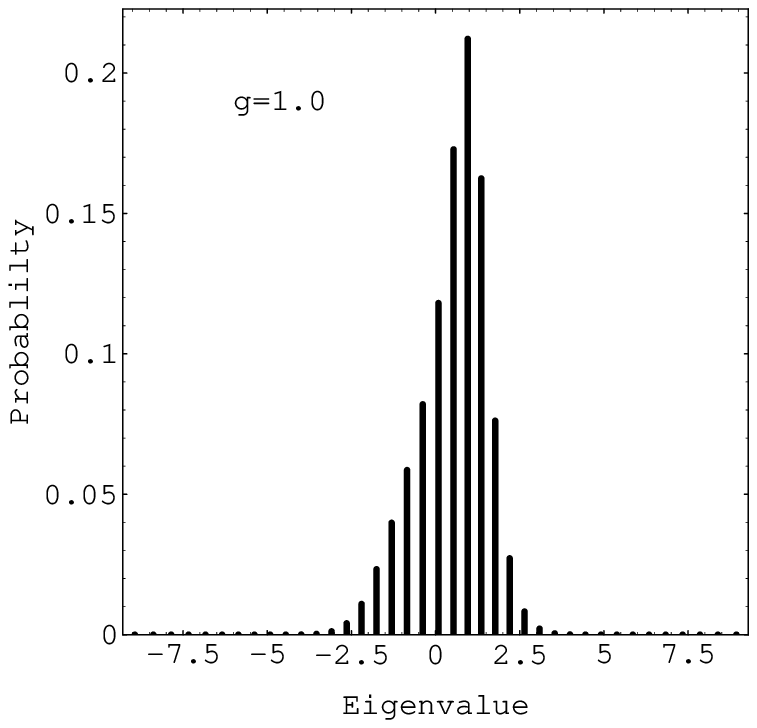 scaled 975}%
\BoxedEPSF{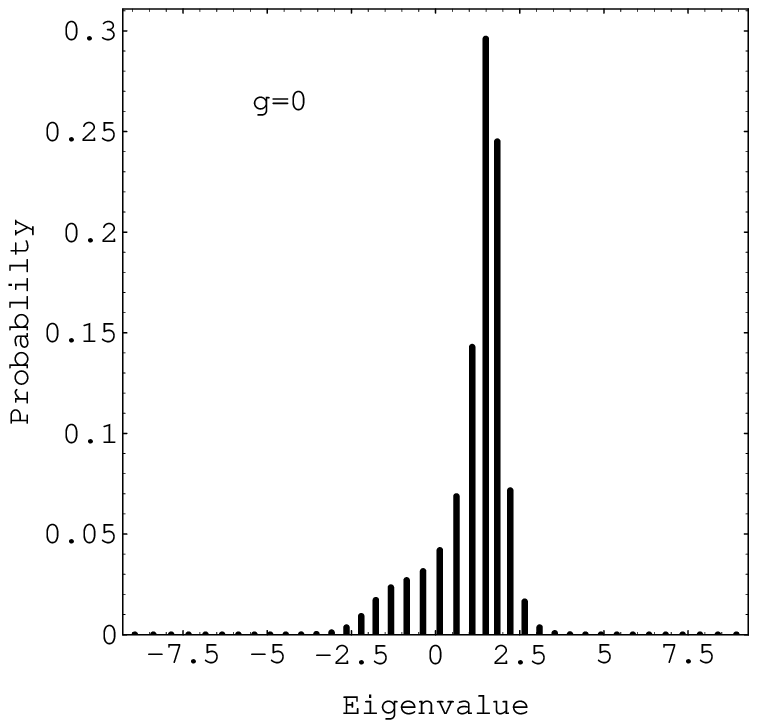 scaled 975}\vspace{0.1in}\\
\hspace*{\fill}(a)\hfill\hfill(b)\hspace*{\fill}\vspace{0.1in}\\
\BoxedEPSF{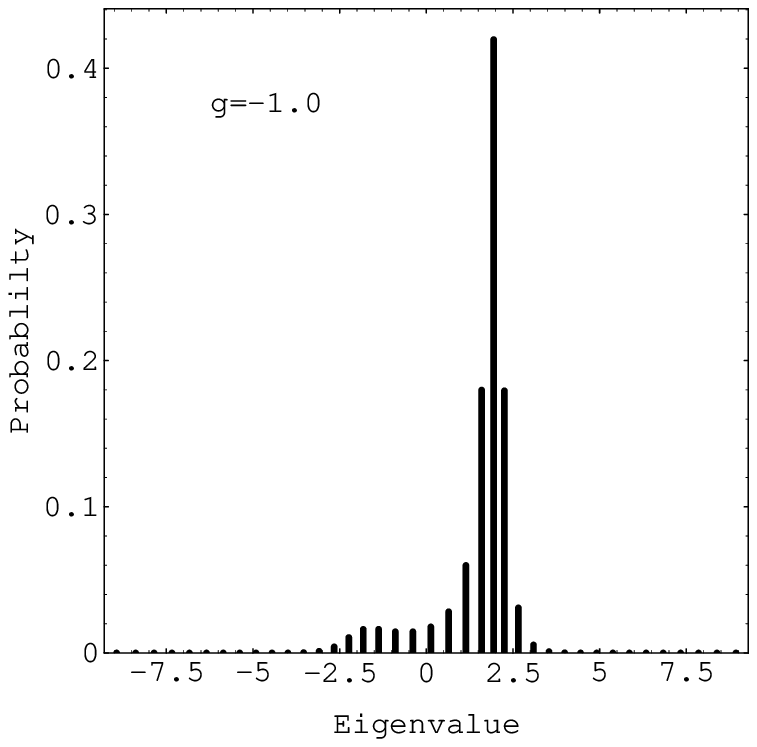 scaled 975}%
\BoxedEPSF{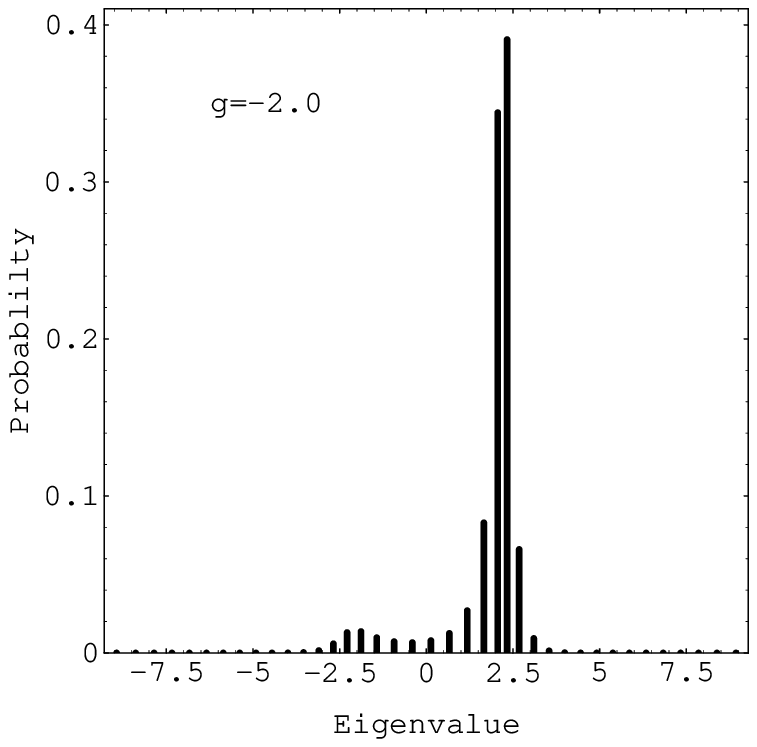 scaled 975}\vspace{0.1in}\\
\hspace*{\fill} (c)\hfill\hfill (d)\hspace*{\fill}
\caption{
\label{ff11} Probability distribution of eigenvalues  of $\protect\sqrt{4 \pi}
\phi$ for the vacuum with couplings (a) $g=1$, (b) $g=0$, (c) $g=-1$, and (d)
$g=-2$. 
$M=1$, $N=50$, and the positive VEV solution to the constraint  equation is used.}
\end{figure}
In general, there are $N+1$ eigenvalues such that $\langle \chi_i | 0 \rangle\neq
0$, independent of $M$.   Thus if one wants to examine the spectrum of the field
operator for the vacuum state, it is better to choose Fock space truncations where
$N$ is large.   With this in mind, one examines the $N=50$ and
$M=1$ case as a  function of $g$ in Figure~\ref{ff11}. Note that near the critical
point, Figure~\ref{ff11}a, the  distribution is approximately equal to the free
field case shown in Figure~\ref{ff10}.  As one moves away from the critical point,
Figures~\ref{ff11}b--d, the distribution  becomes increasingly narrow with a peak
located at the VEV of what would be the minimum of the symmetric double well 
potential in the equal-time paradigm. In addition, there is a small peak
corresponding to minus the VEV.  In the language of the equal-time paradigm, 
there is tunneling between the two minima of the potential. The
spectrum of $\phi$ has been examined for other values of $M$ and
$N$; the results are consistent with the example discussed here.

\subsection{ Physical Picture and Classification of Zero Modes in Gauge Theories } 

When considering a gauge theory, there is a ``zero mode'' problem
associated with the choice of gauge in the compactified case.  This
subtlety, however, is not particular to the light cone; indeed, its
occurrence is quite familiar in equal-time quantization on a
torus \cite{man85,mip94,las92}.  In the present context, the difficulty is that
the zero mode in $A^+$ is in fact gauge-invariant, so that the
light-cone gauge $A^+=0$ cannot be reached.  Thus one has a pair of
interconnected problems: first, a practical choice of gauge; and
second, the presence of constrained zero modes of the gauge field.  In ref.\cite{kar94}
the  generalize gauge fixing in a discrete formalism is described by Kalloniatis and
Robertson.

One defines, for a periodic quantity $f$, its
longitudinal zero mode
\begin {equation}
\langle f \rangle_{o} \equiv {1\over2L}\int_{-L}^L dx^- f(x^-,x_\perp)
\label{zeromode}
\end {equation}
\noindent and the corresponding normal mode part
\begin {equation}
\langle {f} \rangle_{n} \equiv f - \langle {f} \rangle_{o}\; .
\label{normal}
\end {equation}
\noindent The ``global zero mode''---the mode
independent of all the spatial coordinates is denoted by $\langle  {f} \rangle$:
\begin {equation}
\langle  {f} \rangle \equiv {1\over\Omega} \int_{-L}^{L} dx^-   
\int_{-L_\perp}^{L_\perp}d^2x_\perp f(x^-,x_\perp)\; .
\label{ozm}
\end {equation}
\noindent Finally, the quantity which will be of most interest to us is the
``proper zero mode,'' defined by
\begin {equation}
{f_0} \equiv \langle {f} \rangle_{o} - \langle {f}  \rangle\; .
\label{pzmdef}
\end {equation}

By integrating over the appropriate direction(s) of space, one can
project the equations of motion onto the various sectors.  The global
zero mode sector requires some special treatment, and will not be discussed here.
Consider the proper zero mode sector of the equations of motion
\begin {equation}
-\partial_\perp^2{A_0}^+ = g{J_0}^+ 
\label{zm2}
\end {equation}
\begin {equation}
-2(\partial_+)^2{A_0}^+-\partial_\perp^2{A_0}^--2\partial_i\partial_+{A_0}^i = g{J_0}^-
 \label{zm3}
\end {equation}
\begin {equation}
-\partial_\perp^2{A_0}^i+\partial_i\partial_+{A_0}^++\partial_i\partial_j{A_0}^j = g{J_0}^i
\; . 
\label{zm1}
\end {equation}

One first observe that Eq.~(\ref{zm2}), the projection of Gauss' law, is a
constraint which determines the proper zero mode of $A^+$ in terms of
the current $J^+$:
\begin {equation}
{A_0}^+ = -g{1\over\partial_\perp^2}{J_0}^+\; .
\label{zm2solved}
\end {equation}
\noindent The equations \ref{zm3} and \ref{zm1} then determine the
zero modes ${A_0}^-$ and ${A_0}^i$.
Eq.~(\ref{zm2solved}) is clearly incompatible with the strict light-cone
gauge $A^+=0,$ which is most natural in light-cone analyses of gauge
theories.  Here one encounter a common problem in treating axial gauges
on compact spaces.
It is not possible to bring an arbitrary gauge field
configuration to one satisfying $A^+=0$ via a gauge transformation,
and the light-cone gauge is incompatible with the chosen boundary
conditions.  The closest one can come is to set the normal mode part of
$A^+$ to zero, which is equivalent to
\begin {equation}
\partial_-A^+=0\; .
\label{gauge1}
\end {equation}
\noindent This condition does not, however, completely fix the gauge---one is 
free to make arbitrary $x^-$-independent gauge transformations without
undoing Eq.~(\ref{gauge1}).  One may therefore impose further conditions
on $A_\mu$ in the zero mode sector of the theory.

Acting on Eq.~(\ref{zm1}) with $\partial_i$.  The
transverse field ${A_0}^i$ then drops out and one obtain an expression
for the time derivative of ${A_0}^+$:
\begin {equation}
\partial_+{A_0}^+ = g{1\over\partial_\perp^2}\partial_i {J_0}^i\; .
\label{dpap}
\end {equation}
\noindent Inserting this back into Eq.~(\ref{zm1}) one
then find, after some rearrangement,
\begin {equation}
-\partial_\perp^2
\Bigl(\delta^{i}_j - {\partial_i\partial_j \over \partial_\perp^2}\Bigr) {A_0}^j =
g \Bigl(\delta^{i}_j -{\partial_i\partial_j \over \partial_\perp^2}\Bigr){J_0}^j\;.
\label{Atrans}
\end {equation}
Now the operator $(\delta^i_j- \partial_i\partial_j / \partial_\perp^2) $ is nothing
more than the projector of the two-dimensional transverse part of the
vector fields ${A_0}^i$ and ${J_0}^i$.  No trace remains of the
longitudinal projection of the field $(\partial_i\partial_j /
\partial_\perp^2){A_0}^j$ in Eq.~(\ref{Atrans}).  This reflects precisely the
residual gauge freedom with respect to $x^-$-independent
transformations. To determine the longitudinal part, an additional
condition is required.

The general solution to Eq.~(\ref{Atrans}) is
\begin {equation}
{A_0}^i = -g{1\over\partial_\perp^2}{J_0}^i + \partial_i\varphi(x^+,x_\perp)\; ,
\label{secsoln}
\end {equation}
\noindent where $\varphi$ must be independent of $x^-$ but is otherwise
arbitrary.  Imposing a condition on, say, $\partial_i{A_0}^i$ will
uniquely determine $\varphi$.
In ref.\cite{kap94}, for example, the condition $\partial_i{A_0}^i=0$
was proposed as being particularly
natural. This choice, taken with the other gauge conditions has been
called the ``compactification gauge.''  In this case
\begin {equation}
\varphi = g{1\over(\partial_\perp^2)^2}\partial_i {J_0}^i\; .
\label{compgp}
\end {equation}
Of course, other choices are also possible.  For example, one can generalize
Eq.~(\ref{compgp}) to
\begin {equation}
\varphi = \alpha g{1\over(\partial_\perp^2)^2}\partial_i{J_0}^i\; ,
\label{gencompgp} 
\end {equation}
with $\alpha$ a real parameter.  Then the ``generalized compactification gauge.''
condition corresponding to this solution is
\begin {equation}
\partial_i {A_0}^i=-g(1-\alpha){1\over\partial_\perp^2}\partial_i {J_0}^i\; .
\label{zmgaugecond}
\end {equation}
An arbitrary gauge field configuration $B^\mu $ can be brought to one
satisfying Eq.~(\ref{zmgaugecond}) via the gauge function
\begin {equation}
\Lambda(x_\perp) =- {1\over\partial_\perp^2} 
\Bigl[ g (1 - \alpha) {1\over\partial_\perp^2} \partial_i {J_0}^i
+ \partial_i {B_0}^i \Bigr]\; .
\label{gaugetrans}
\end {equation}
This is somewhat unusual in that $\Lambda(x_\perp)$ involves the
sources as well as the initial field configuration, but this is
perfectly acceptable.  More generally, $\varphi$ can be any
(dimensionless) function of gauge invariants constructed from the
fields in the theory, including the currents $J^\pm$. For our purposes
Eq.~(\ref{zmgaugecond}) suffices.

One now has relations defining the proper zero modes of $A^i$,
\begin {equation}
{A_0}^i = -g{1\over\partial_\perp^2}\Bigl(
\delta^i_j-\alpha{\partial_i\partial_j\over\partial_\perp^2}\Bigr) {J_0}^j\; ,
\label{aizeromode}
\end {equation}
\noindent as well as ${A_0}^+$ Eq.~(\ref{zm2solved}).  All that remains is to use the
final constraint Eq.~(\ref{zm3}) to determine ${A_0}^-.$  Using eqs.
\ref{dpap} and \ref{zmgaugecond}, one finds that Eq.~(\ref{zm3}) can be written as
\begin {equation}
\partial_\perp^2{A_0}^- = -g{J_0}^--2\alpha g{1\over\partial_\perp^2}
\partial_+\partial_i {J_0}^i\; .
\label{solvingam0}
\end {equation}
After using the equations of motion to express $\partial_+{J_0}^i$ in
terms of the dynamical fields at $x^+=0$, this may be
straightforwardly solved for ${A_0}^-$ by inverting the
$\partial_\perp^2.$ In what follows, however, one has no need of
${A_0}^-.$ It does not enter the Hamiltonian, for example; as usual,
it plays the role of a multiplier to Gauss' law Eq.~(\ref{zm1}), which one is
able to implement as an operator identity.
                               
The extension of the present work to the case of QCD is complicated by
the fact that the constraint relations for the gluonic zero modes are
nonlinear, as in the $\phi^4$ theory.  A perturbative solution of the
constraints is of course still possible, but in this case, since the
effective coupling at the relevant (hadronic) scale is large, it is
clearly desirable to go beyond perturbation theory.  In addition,
because of the central role played by gauge fixing in the present
work, one may expect complications due to the Gribov
ambiguity\cite{gri78}, which prevents the selection of unique
representatives on gauge orbits in nonperturbative treatments of
Yang-Mills theory.  Preliminary step in this direction on the pure glue
theory in 2+1 dimensions is found in ref.\cite{kap94}. There one finds that some of
the nonperturbative techniques used recently in 1+1 dimensions \cite{bep93,piv94} can be
applied.

\subsection{Dynamical Zero Modes}
 
Our concern in this section is with those zero modes that are
true dynamical independent fields. They can arise  
due to the boundary conditions in gauge theory preventing one from fully implement the 
traditional light-cone gauge $A^{+}=0$. The development of the
understanding of this problem in DLCQ can be traced in Refs.
\cite{mcc88,hek91b,kap93b,kap94}.  It has its analogue
in instant form approaches to gauge theory \cite{man85,het93}. 

Consider the zero mode {\it subsector} of
the pure glue theory in (1+1) dimension, namely where only zero mode external sources
excite only zero mode gluons. This is not an approximation
but rather a consistent solution, a sub-regime within the complete
theory. A similar framing of the problem lies behind the work of L\"uscher 
\cite{lus83}
and van Baal \cite{van92} using the instant form Hamiltonian approach to pure glue
gauge theory in 3+1 dimensions. The beauty of this reduction in the 1+1 dimensional
theory is two-fold. First, it yields a theory which is exactly soluble. This is useful
given the dearth of soluble models in field theory.  Secondly, the zero mode theory
represents a paring down to the point where the front and instant forms are manifestly
{\it identical}, which is nice to know indeed.  

Consider an SU(2) non-Abelian gauge theory in 1+1 dimensions with classical sources 
coupled to the gluons. The Lagrangian density is
\begin{equation}
{\cal L} = {1\over 2}\, {\rm Tr}\, (F_{\mu \nu} F^{\mu \nu}) + 2\, {\rm Tr}\, (J_\mu A^\mu) 
\end{equation}
where $F_{\mu \nu} = \partial _{\nu} A_{\nu} - \partial_{\nu} A_{\mu} -g[A_{\mu}, 
A_{\nu}]$. With a finite interval in $x^-$ from $-L$ to $L$, one imposes periodic
boundary conditions  on all gauge potentials $A_\mu$.

One cannot eliminate the zero mode of the gauge potential. The
reason is evident: it is {\it invariant} under periodic gauge
transformations. But of course one can always perform a rotation in color space.  
In line with other authors \cite{anp90,prf89,frn81}, one chooses this so that
$\stackrel{o}{A_3^+}$  is the only non-zero element, since in our representation only
$\sigma^3$ is diagonal.  In addition, one can impose the subsidiary gauge condition
$ \stackrel{o}{A^-_3} = 0$ The reason is that there still remains freedom to perform
gauge transformations that depend only on light-cone time $x^+$ and the color matrix
$\sigma^3$.

The above procedure would appear to have enabled complete fixing of the gauge. 
This is still not so. Gauge transformations
\begin{equation}
V = \exp\{i x^- ({{n\pi} \over {2L}}) {\sigma}^3\}
\label{GribU}
\end{equation}
\noindent generate shifts, according to Eq.(\ref{gaugetrans}), in the zero mode
component
\begin{equation}
\stackrel{o}{A^+_3} \rightarrow \stackrel{o}{A^+_3} + {{n\pi}\over{gL}}
\;.
\end{equation}
All of these possibilities, labeled by the integer $n$, of course still satisfy
$\partial_- A^+=0$, but as one sees $n=0$ should not really be included. One can verify
that the transformations $V$ also preserve the subsidiary condition.   One notes that the
transformation is 
$x^-$-dependent and {\it $Z_2$ periodic}. It is
thus a simple example of a Gribov copy \cite{gri78} in 1+1 dimensions.   
Following the conventional procedure one demands 
\begin{equation}
\stackrel{o}{A^+_3} \neq {n \pi \over gL}\;, \quad  n= \pm1, \pm2, \ldots 
\;.
\end{equation}
This eliminates singular points at the Gribov `horizons' which
in turn correspond to a vanishing Faddeev-Popov determinant
\cite{van92}.

For convenience we henceforth use the notation
\begin{equation} 
\stackrel{o}{A^+_3} = v \;, \quad  
 x^+ = t \;, \quad  
w^2  =  { {\stackrel{o}{J^+_+} \stackrel{o}{J^+_-}} \over {g^2} } \quad {\rm{and}} \quad  
\stackrel{o}{J^-_3} = {B\over2}  
\;. 
\end{equation}
The only conjugate momentum is 
\begin{equation} 
p\, \equiv\, \stackrel{o}{\Pi^-_3} \, = \, \partial^- \!\! \stackrel{o}{A^+_3}\,  =\, \partial^- v  
\;.
\end{equation}
The Hamiltonian density  $ T^{+ -}\, = \, \partial^- \!\! \stackrel{o}{A^+_3} \Pi^-_3 -
\cal L $  leads to the Hamiltonian 
\begin{equation}
 H =  {1 \over 2} [ {p}^2  +  {w^2 \over v^2} + B v] (2L)  
\;.
\label{qmHamil} 
\end{equation}
Quantization is achieved by imposing a commutation relation at equal light-cone time 
on the dynamical degree of freedom. Introducing the variable 
$ q = 2L v $, 
the appropriate commutation relation is    $[q(x^+), p(x^+)] = i .$  
The field theoretic problem reduces to quantum mechanics of a 
single particle as in Manton's treatment of the Schwinger model in
Ref.\cite{man85}.  One thus has to solve the Schr\"odinger equation 
\begin{equation}
{1 \over 2} (- {d^2 \over dq^2} + {(2Lw)^2 \over q^2} + {{B q} \over {2L}})\psi = 
{\cal E}  \psi, 
\label{schrodeq} 
\end{equation}
\noindent with the eigenvalue ${\cal E} = E/(2L)$ actually being an energy density. 

All eigenstates $\psi$ have the quantum numbers of the naive vacuum adopted in
standard front form field theory: all of them
are eigenstates of the light-cone momentum operator $P^+$ with zero eigenvalue. 
The true vacuum is now that state with lowest $P^-$ eigenvalue. 
In order to get an exactly soluble system one eliminates the source $2B =
\stackrel{o}{J^-_3}$.  

The boundary
condition that is to be imposed comes from the treatment of the Gribov problem. 
Since the wave function vanishes at $q=0$ one must
demand that the wavefunctions vanish at the first Gribov horizon $q = \pm 2\pi / g$. 
The overall constant $R$ is then fixed by normalization. This leads to   
the energy density only assuming the discrete values
\begin{equation}
 {\cal E}_m^{(\nu)} = {g^2 \over {8 {\pi}^2}} (X_m^{(\nu)})^2 
, \quad   m = 1,2, \dots, 
\label{Evals}
\end{equation}
\noindent where $X_m^{(\nu)}$ denotes the m-th zero of the $\nu$-th Bessel function
$J_\nu$. In general, these zeroes can only be obtained numerically. 
Thus
\begin{equation}   
\psi_m (q) = R\sqrt{q} J_{\nu} (\sqrt{2 {\cal E}_m^{(\nu)}} q)  
\end{equation}
is the complete solution. The true vacuum is the state of lowest energy namely with
$m=1$.

The exact solution is genuinely non-perturbative in character. It describes
vacuum-like states since for all of these states $P^+=0$. Consequently, they all have
zero invariant mass $M^2 = P^+ P^-$. The states are labeled
by the eigenvalues of the operator $P^-$. The linear dependence on $L$ in the result
for the discrete {\it energy} levels is also consistent with what one would
expect from a loop of color flux running around the cylinder. 

In the source-free equal time case Hetrick \cite{het93} uses a wave function that is
symmetric about $q = 0$. For our problem this corresponds to
\begin{equation}
\psi_m(q) = N \cos (\sqrt{2 \epsilon_m} q )
\;.
\end{equation}
where N is fixed by normalization. At the boundary of the fundamental
modular region
$q =2\pi /g$ and $\psi_m = (-1)^m N $, thus $\sqrt{2 \epsilon_m} 2 \pi /g =m \pi$ and
\begin{equation}
\epsilon = {g^2 (m^2-1)\over{8}}
\;.
\end{equation}
Note that $ m=1 $ is the lowest energy state and has as expected one node in the
allowed region $ 0\le g \le 2 \pi /g $. Hetrick \cite{het93} discusses the connection
to the results of Rajeev \cite{raj92} but it amounts to a shift in
$\epsilon$ and a redefining of $m \rightarrow m/2$. It has been argued
by van Baal that the correct boundary condition at $q=0$ is $\psi(0) = 0 $. This would
give a sine which matches smoothly with the Bessel function solution. This calculation
offers the lesson that even in a front form approach, the vacuum might not be just the
simple Fock vacuum. Dynamical zero modes do imbue the vacuum with a rich structure.

\section{Acknowledgments}

It is a pleasure to thank the organizers of {\sc Orbis Scientiae 1996}
for providing such a stimulating and enjoyable atmosphere.  This work
was supported in part by a grant from the the US Department of Energy.

\end{document}